\documentclass[14pt]{extarticle}
\usepackage[russian]{babel}
\usepackage{inputenc}
\usepackage{amsmath,latexsym,amsthm}
\usepackage{graphics}
\usepackage{graphicx}
\usepackage{amssymb}
\usepackage{color}
\usepackage[colorlinks]{hyperref}

\usepackage{enumitem}

\usepackage{amsmath}

\newcommand{\bfbeta}{\mbox{\boldmath $\beta$}}

\newcommand{\bfk}{{\bf k}}

		\newcommand{\bfx}{{\bf x}}	\newcommand{\bfX}{{\bf X}}
\newcommand{\bfy}{{\bf y}}

\begin{document}
\begin{center}
\textbf{\large Iterative variable selection for high-dimensional data with binary outcomes}\\
\textbf{Nilotpal Sanyal $^{1}$ }\\  
\textsc{{$^{1}$ Indian Statistical Institute}}\\  
\textit{{nilotpal.sanyal@gmail.com}}\\
\end{center}


\makeatletter
\renewcommand{\@evenhead}{\vbox{\thepage \hfil {\it Nilotpal Sanyal }   \hrule }}

\renewcommand{\@oddhead}{\vbox{\hfill
{\it \small{VI International conference "STATISTICS and its Applications"\,, 2022 y., Namangan}
 }\hfill \thepage \hrule}} \makeatother
\label{firstpage}

\sloppy


\section*{Introduction}
High-dimensional data with binary outcomes are common in biological sciences and related fields such as omics, epidemiology, and healthcare, environmental sciences, physical and engineering sciences, and business and policy making. Often, such data are sparse so that only a small number of all available variables truly affect the binary outcome. Several frequentist and Bayesian methods exist in the literature that provide a high-dimensional variable selection for binary outcomes (see [1]). Recently, [2] proposed a high-dimensional variable selection scheme for continuous outcomes in the context of genome-wide association studies. They introduce the concept of a `structured screen-and-select' strategy and examined the use of non-local priors within the same. Here, we build upon their work and propose an iterative variable selection scheme for high-dimensional data with binary outcomes.

\section*{Methods}
We consider the problem of variable selection with binary outcomes for $n$ subjects and $p$ independent variables where $p$ can be much larger than $n$ ($p\gg n$). Let $\bfy^{n \times 1} = (y_1,y_2,\ldots,y_n)$ denote the vector of binary outcomes and $\bfX^{n \times p}$ denote the design matrix whose columns correspond to the independent variables. 

We propose an iterative scheme for variable selection that uses non-local priors. In each iteration, a smaller number of variables are screened from all candidate variables using a measure of association, and non-local prior-based Bayesian model selection is performed with the screened variables. We call these two steps respectively the \emph{screening step} and the \emph{selection step} of an iteration. First, we describe the data distribution and the priors used for the selection steps and afterward, describe the full iterative scheme.

\subsection*{Data distribution and priors}
The selection steps perform Bayesian model selection using non-local priors. Here, a model consists of a collection of variables. Let us denote by $p_s$ the number of screened variables undergoing the $s$th selection step. The number of all possible models at the $s$th selection step is $2^{p_s}$. Suppose for model $\bfk$, $\bfk=1,\ldots,2^{p_s}$, $\bfX_{\bfk}^s$ denotes the design matrix whose $i$th row, $\bfx_{i\bfk}^s$, corresponds to the $i$th subject and $\bfbeta_{\bfk}^s=(\beta_{1,\bfk}^2,\ldots,\beta_{|\bfk|,\bfk}^2)$ denotes the vector of regression coefficients where $|\bfk|$ is the number of variables in model $\bfk$. For any $\bfk$, given $\bfx_{i\bfk}^s$, we assume a logistic regression model for the binary phenotype $y_i$, i.e.,
$$
Pr(y_i=1 | \bfx_{i\bfk}^s) = \frac{\exp({\bfx_{i\bfk}^s}^T \bfbeta_{\bfk}^s)}{1+\exp({\bfx_{i\bfk}^s}^T \bfbeta_{\bfk}^s)} . \eqno(1)
$$

Model selection involves two types of uncertainties--the uncertainty regarding selecting a model out of all possible models forming the model space, and the uncertainty regarding the estimation of the parameters of a selected model from the parameter space. In the Bayesian context, one needs to specify prior distributions over both the model space and the parameter space.

We consider two different choices for non-local priors over the parameter space---the product moment prior (pMOM prior) and the product inverse moment prior (piMOM prior) [3]. The pMOM prior is the product of individual moment (MOM) priors for the regression parameters $\bfbeta_{\bfk}^s$ and is given by 
$$
\pi( \bfbeta_\bfk^s \; | \; r, \tau_1 )   \; = \;   M_{|\bfk|}^{-1}   \;  (\phi\tau_1)^{- |\bfk|/2 - r|\bfk|}   \;   \prod_{j=1}^{|\bfk|}  {\beta_{j,\bfk}^s}^{2r}  \;   
\exp\left[ - \frac{1}{2\phi\tau_1} \sum_{j=1}^{|\bfk|} {\beta_{j,\bfk}^s}^2   \right],   \eqno(2)
$$
where $\tau_1$ is the common scale parameter and $r$ is the common order of the MOM priors, $\phi$ is a dispersion parameter and $M_{|\bfk|}$ is a marginalizing constant given by $M_{|\bfk|}  =  (2\pi)^{- |\bfk|/2}  \left\{ \prod_{l=1}^{r} (2l-1) \right\}^{|\bfk|} $.

The piMOM prior is the product of individual inverse moment (iMOM) priors for the regression parameters $\bfbeta_{\bfk}^s$ and is given by 
$$
\pi( \bfbeta_\bfk^s \; | \; \nu, \tau_2 )  \; = \;   \frac{ (\phi\tau_2)^{\nu|\bfk|/2} }{ (\Gamma( {\nu \over 2} ))^{|\bfk|} }  \; \prod_{j=1}^{|\bfk|} |\beta_{j,\bfk}^s|^{-(\nu+1)}  \; \exp \left( - \phi\tau_2 \sum_{j=1}^{|\bfk|} {1 \over {\beta_{j,\bfk}^s}^2}   \right), \eqno(3)
$$
where $\tau_2$ is the common dispersion parameter and $\nu$ is the common shape parameter of the iMOM priors.

For the model space, we assume a beta-binomial prior [3], given by $\pi( \bfk | \gamma ) = \gamma^{|\bfk|} \; (1-\gamma)^{ p_s - |\bfk|} \text{ where }  \gamma \sim beta(1,1)$ distribution. In the Bayesian paradigm, the posterior probability of each model is computed and the model with the highest posterior probability is generally considered as the selected model. 

\subsection*{The proposed iterative scheme}
The proposed iterative variable selection scheme uses the structured screen-and-select strategy [2] whereby in each iteration, smaller subsets of candidate variables are screened using measures of association between the outcome and the independent variables, and among the independent variables. Screened variables are evaluated for being included in the final selection based on non-local prior-based Bayesian model selection. The scheme is described below.

\begin{itemize}[topsep=10pt]
\item In the first iteration ($s=1$), the association of all candidate variables with the binary outcome is ascertained in terms of maximum marginal likelihood estimates (MMLE) [4]. For the $j$th candidate variable, the MMLE is given by
$$
\widehat{\beta_{j}^1}_{mmle} = \text{the last element of }\operatorname*{arg\,min}_{\beta0,\beta_j^1} \sum_{i=1}^n L(y_i,\beta_0 + x_{ij}^1 \beta_j^1),
$$
where $L$ is the usual logarithmic loss function for the logistic regression of $y_i$ on independent variable $x_j$ and an intercept. The $k_0$ variables that have the largest absolute values of the association are called the \emph{leading variables}. For each leading variable, we collect all candidate variables that have an absolute Pearson correlation coefficient value $\geq r$ with that leading variable where $r$ is a given threshold; this collection is called the \emph{leading set}. For each leading set, non-local prior-based model selection is performed according to the method in [3], using the data distribution and non-local priors described in Equations 1-3. The variables contained in the higher posterior probability model of each leading set are included in the final selection. The remaining variables of each leading set are excluded from subsequent analysis.

\item Subsequent iterations, $s=2,3,\ldots$, proceed as before except for the evaluation of the association between candidate variables and the outcome which is done as follows. Suppose $\bfX_{sel}^{(s-1)}$ is the set of variables selected in iterations 1 through $(s-1)$, and $\bfx_{i,sel}^{(s-1)}$ is its $i$th row. In iteration $s$, the association of the remaining candidate variables with the binary outcome is ascertained in terms of their MMLEs in presence of the variables $\bfX_{sel}^{(s-1)}$ in the model. For the $j$th candidate variable of iteration $s$, the MMLE is given by
$$
\widehat{\beta_{j}^s}_{mmle} = \text{the last element of }\operatorname*{arg\,min}_{\beta_0,\bfbeta_{sel}^{(s-1)},\beta_j^s} \sum_{i=1}^n L(y_i,\beta_0+\bfx_{i,sel}^{(s-1)} \bfbeta_{sel}^{(s-1)} + x_{ij}^s \beta_j^s),
$$
where $\bfbeta_{sel}^{(s-1)}$ are the regression coefficients for variables $\bfx_{i,sel}^{(s-1)}$. 

\item Variables are selected through this structured screen-and-select strategy until we reach a desired number, $m$, of selected variables, or until the number of iterations that select no variables reaches a maximum allowed value, $maxno$. 
\end{itemize}

The proposed scheme has four tuning parameters $k_0$, $r$, $m$ and $maxno$ that determine the number of selected variables. In the following application, we set them using the heuristic guidelines provided in [2].

\section*{Application}
We used the proposed method to select single nucleotide polymorphisms (SNPs) for binary phenotypes. The genotype data were simulated using methods described in [2] such that the linkage disequilibrium structure of the genotype matrix resembled that of real genotyped SNPs in chromosome 1 of the human genome. The data contain genotypes for $p=20000$ SNPs for $n=2000$ subjects. 20 SNPs were randomly chosen as causal SNPs and their standardized effect sizes were randomly generated from the N(0,1) distribution. The phenotype data were generated according to the model in Equation~1.

We analyze the data using our proposed method as well as LASSO, elastic net ($\alpha=0.75,0.5,0.25$), and ISIS with SCAD penalty (see references within [1] and [2]). For our method, we set $\phi=1$, $\tau_1=\tau_2=0.2$, $k_0=1$, $r=0.3$ and $maxno=3$. For the other methods, default cross-validation tuning was used. The performance of all the methods was compared in terms of true positive rate (TPR) and false discovery rate (FDR), and shown in the following figure. Our proposed method has evidently shown a very competitive profile compared to the best of the alternatives.

\begin{figure}
\includegraphics[width=\textwidth]{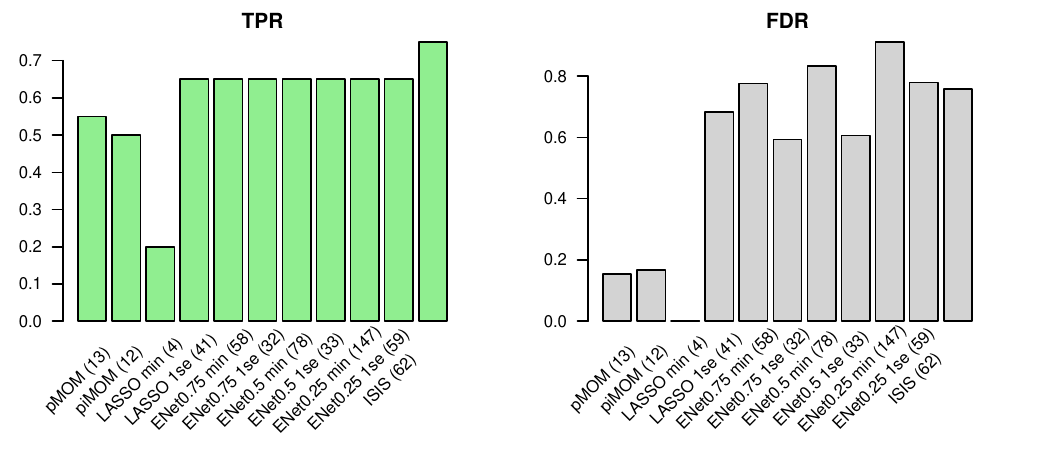}
\caption{Comparison of the proposed method (pMOM and piMOM) with LASSO, Elastic net, and ISIS in terms of true positive rate and false discovery rate based on genomic SNP data with $p=20000$ SNPs and $n=2000$ subjects and 20 truly causal SNPs. The labels in the x-axis show the names of the various methods and the number of selected variables within parentheses.}
\end{figure}

\section*{Conclusion}
We propose a competitive variable selection method for high-dimensional data with binary outcomes. The proposed method has been implemented within the existing R package GWASinlps. In this work, we have considered setting the tuning parameters based on heuristics. In the future, we wish to develop objective criteria for optimally estimating these tuning parameters from the data.


\clearpage

\textcolor{red}{Annotation:}\\

\textbf{Nilotpal Sanyal (Kolkata, India)}  

\textbf{High-dimensional variable selection for binary data}

\textbf{Abstract.} \textit{We propose an iterative variable selection scheme for high-dimensional data with binary outcomes. The scheme adopts a structured screen-and-select framework and uses non-local prior-based Bayesian model selection within the same. The structured screening is based on the association of the independent variables with the outcome which is measured in terms of the maximum marginal likelihood estimator. Performance comparison with several well-known methods in terms of true positive rate and false discovery rate shows that our proposed method stands to be a competitive alternative for sparse high-dimensional variable selection with binary outcomes. The method has been implemented within the R package GWASinlps.
}

\textbf{Key words:}
\emph{High-dimensional, variable selection, nonlocal prior}


\medskip
\begin{center}
\textbf{REFERENCE}
\end{center}
\begin{enumerate}


\bibitem{1} {Nikooienejad, A., Wang, W., \& Johnson, V. E.}, {Bayesian variable selection for binary outcomes in high-dimensional genomic studies using non-local priors. Bioinformatics, 2016, N. 32(9), p.~1338--1345. }
\bibitem{2} {Sanyal, N., Lo, M. T., Kauppi, K., Djurovic, S., Andreassen, O. A., Johnson, V. E., \& Chen, C. H.}, {GWASinlps: non-local prior based iterative SNP selection tool for genome-wide association studies. Bioinformatics, 2019, N. 35(1), p.~1--11. }
\bibitem{3} {Johnson, V. E., \& Rossell, D.}, {Bayesian model selection in high-dimensional settings. Journal of the American Statistical Association, 2012, N. 107(498), p.~649--660. }
\bibitem{4} {Fan, J., Samworth, R., \& Wu, Y.}, {Ultrahigh dimensional feature selection: beyond the linear model. The Journal of Machine Learning Research, 2009, N. 10, p.~2013--2038. }



\end{enumerate}
\end{document}